\documentclass[letterpaper,12pt]{article}
\usepackage{amssymb}   %% LaTeX 2e (preferred)
\usepackage{osajnl2} %% do not use with REVTeX4
\usepackage[draft]{hyperref} %% optional
\usepackage{subfigure}
\usepackage{graphicx}

\usepackage{graphics}

\begin{document}

\title{Control of two-atom entanglement with two thermal fields in coupled cavities}

%% For REVTeX it is possible to automate superscript and e-mail callouts with the superscriptaddress option; see REVTeX4 documentation.

\author{Li-Tuo Shen,$^{1}$ Zhen-Biao Yang,$^{2}$ Huai-Zhi Wu,$^{1}$ Xin-Yu Chen,$^{1}$ and Shi-Biao Zheng$^{1,\dagger }$}
\address{$^1$Lab of Quantum Optics, Department of Physics,\\
Fuzhou University, Fuzhou 350002, China}
\address{$^2$Key Laboratory of
Quantum Information, \\University of Science and Technology of
China, Chinese Academy of Sciences, Hefei 230026, China}
\address{$^\dagger $Corresponding author: sbzheng11@163.com}

\begin{abstract}
The dynamical evolution of a quantum system composed of two coupled
cavities, each containing a two-level atom and a single-mode thermal
field, is investigated under different conditions. The entanglement
between the two atoms is controlled by the hopping strength and the
detuning between the atomic transition and the cavities. We find
that when the atomic transition is far off-resonant with both the
eigenmodes of the coupled cavity system, the maximally entangled
state for the two atoms can be generated with the initial state in
which one atom is in the ground state and the other is in the
excited state. When both the two atoms are initially in the excited
state, the entanglement exhibits period sudden birth and death. By
choosing appropriate parameter values, the initial maximal
entanglement of the two atoms can be frozen. The relation between
the concurrence and cooperative parameter is calculated.
\end{abstract}

\ocis{270.2500, 270.5580}% REPLACE WITH CORRECT OCIS CODES FOR YOUR ARTICLE
                          % NOTE: \ocis{} IS ALIASED TO \pacs{} BUT MUST
                          % FORMAT THE TERMS CORRECTLY FOR EACH JOURNAL

\maketitle %% null function with osajnl.sty

\section{Introduction}

Quantum system inevitably interacts with the surrounding heat
environments, i.e., thermal fields. The thermal field, which is
emitted from a source in thermal equilibrium, has become an
important topic in quantum theory of damping and quantum experiment
of laser \cite{Scully-Quantum-Optics}. Under the condition of
constant average energy, the thermal field maximizes the entropy of
the field and is always considered to be chaotic
\cite{PRA-54-804-1996}. However, Bose \emph{et al.}
\cite{PRL-87-050401-2001} showed that a single two-level atom and a
thermal field could be entangled due to their linear interaction.
Recently, the model in which two atoms interact with a single-mode
thermal field has attracted great interest
\cite{PRA-65-040101-2002,OC-283-4676-2010,JOSAB-28-2087-2011,
JOB-4-425-2002,JPA-37-5477-2004,JOB-7-769-2005,JPB-39-3805-2006}.
Kim \emph{et al.} \cite{PRA-65-040101-2002} considered two identical
two-level atoms which are resonantly coupled to a single-mode
thermal field and found the field could entangle two atoms which are
initially in a separate state, while Zhang \cite{OC-283-4676-2010}
generalized Kim's idea to the case that two atoms are slightly
detuned from the thermal field. El-Orany \cite{JOSAB-28-2087-2011}
gave a exact treatment for entanglement between two atoms when they
interact simultaneously with a single-mode thermal field through
multiphoton exchange. On the other hand, the fundamental dissipative
process of the radiation field inside a cavity may be understood
from a simple model where the mode of interest is coupled to a
multi-mode thermal reservoir. There have been some studies devoted
to the influence of thermal noise on the behavior of different
quantum system
\cite{PRA-57-1451-1998,PRL-88-197901-2002,JOB-6-378-2004,IJMPB-2011-25-2681,arXiv:1111.5208v1,PS-2012-T147-014022}
and to the possibility of the quantum information processing via
thermal fields
\cite{PRA-66-060303-2002,PRA-68-035801-2003,PRA-75-032114-2007}.
However, previous theoretical schemes and experiments focus on the
case in which two atoms interact with a common single-mode thermal
field and the maximal entanglement can be obtained only when the
atomic transition frequency is highly detuned from the field
frequency.

Recently, considerable attentions have been paid to the
coupled-cavity system
\cite{PRA2008-78-063805,LPR2008-2-527,PRL2006-96-010503,EPJD2011-61-737,
PRA2007-75-012324,EPJD2008-50-91,PRA2010-82-012307}, which is
promising to overcome the difficulty of individual addressability
existing in a single cavity. Dynamics in a coupled-cavity array
offers a basic setting for distributed quantum information
processing and can be studied for a wide range of parameters, such
as atom-cavity detuning, intercavity coupling rate, etc. In this
paper, we investigate the entanglement dynamics for two atoms
trapped in two coupled cavities, each containing one single-mode
thermal field.

Contrary to previous studies in which two atoms simultaneously
interact with only one single-mode thermal field, here we consider a
quantum system comprising two identical two-level atoms trapped in
two coupled cavities, each of which is initially in a thermal state.
The entanglement dynamics of the atomic system depends upon the
photon hopping strength, the detuning between atomic transition
frequency and field frequency, and the initial state. We find that
even when the atoms symmetrically and resonantly interact with the
local thermal fields, the maximally entangled state for the atoms
can be generated with the initial state in which one atom is in the
ground state and the other is in the excited state. This result is
obviously different from the case that two atoms symmetrically and
resonantly interact with a common thermal field
\cite{PRA-65-040101-2002}, in which the obtainable entanglement is
rather low. Under certain conditions, the thermal fields can induce
the periodical entanglement sudden birth and death when the atoms
are both initially in excited states or ground states. Remarkably,
the peaks of the concurrence do not decrease monotonically when the
evolution time increases, which is distinguished from previous
examples
\cite{OL-2008-33-270,JPB-39-S621-2006,PRA-77-054301-2008,PRL-101-080503-2008}.
By choosing appropriate photon hopping strength and detuning between
the atomic transition and the cavities, one can freeze the
entanglement between the two atoms when they are initially in the
maximally entangled state, which may have practical application in
quantum memory and quantum storage
\cite{Nature-473-190-2011,PRA-84-010301-2011,RMP-82-1041-2010}. The
influences of atomic spontaneous emission and photon leakage out of
the cavity are analyzed.

\section{The Theoretical Model}
 The coupled-cavity system under
consideration is shown in Fig. 1. Two identical two-level atoms $1$
and $2$ are trapped in two coupled cavities, each containing a
single-mode thermal field $\rho_{f_{i}}$ ($i$ $=$ $1$, $2$). In the
frame rotating at the cavity frequency, the total Hamiltonian under
the rotating-wave approximation is ($\hbar$ $=$ $1$):
\begin{eqnarray}\label{e1}
&H=\sum_{i=1}^{2}\bigg[\delta_{i}a_{i}^{\dagger}a_{i}+g_{i}\big(S_{i}^{+}a_{i}+S_{i}^{-}a_{i}^{\dagger}\big)\bigg]
+J\big(a_{1}^{\dag}a_{2}+a_{1}a_{2}^{\dag}\big),&
\end{eqnarray}
where $S_{i}^{+}$ $=$ $|e_{i}\rangle\langle g_{i}|$ and $S_{i}^{-}$
$=$ $|g_{i}\rangle\langle e_{i}|$ with $|e_{i}\rangle$ and
$|g_{i}\rangle$ being the excited state and ground state of the
$i$th atom. $a_{i}^{\dag}$ and $a_{i}$ are the creation and
annihilation operators for the $i$th thermal field, $g_{i}$
describes the coupling strength between the $i$th atom and the $i$th
thermal field, and $\delta_{i}$ is the detuning between the atomic
transition and the local cavity mode. $J$ represents the coherent
photon hopping strength between the two cavities.

The density operator of the $i$th thermal radiation field is
described by
\begin{eqnarray}\label{e2-e3}
&\rho_{f_{i}}=\sum_{n_{i}=0}^{\infty}P_{i}(n_{i})|n_{i}\rangle\langle
n_{i}|,&\\\cr
&P_{i}(n_{i})=\frac{\bar{n}_{i}^{n_{i}}}{(\bar{n}_{i}+1)^{(n_{i}+1)}},&
\end{eqnarray}
where $\bar{n}_{i}$ $=$ $(e^{\hbar w_{f_i}/k_{B}T_{i}}-1)^{-1}$ is
the mean photon number and $|n_{i}\rangle$ is the Fock state.
$k_{B}$ is the Boltzmann constant, $w_{f_i}$ is the frequency of the
cavity mode, and $T_{i}$ is the temperature.

The excitation number of the total system is conserved during the
state evolution since the excitation number operator
$\hat{N}=\sum_{i=1}^{2}(|e_{i}\rangle\langle
e_{i}|+a_{i}^{\dagger}a_{i})$ commutes with the Hamiltonian $H$. For
different initial excitation numbers, the system will evolve in
different invariant subspaces. The zero-excitation subspace is:
\begin{eqnarray}\label{e4}
\Gamma_{0}=\big\{|g_{1}g_{2}\rangle|0\rangle_{1}|0\rangle_{2}\big\},
\end{eqnarray}
and the single-excitation subspace is:
\begin{eqnarray}\label{e5}
\Gamma_{1}&=&\big\{|e_{1}g_{2}\rangle|0\rangle_{1}|0\rangle_{2},|g_{1}e_{2}\rangle|0\rangle_{1}|0\rangle_{2},
\cr&&|g_{1}g_{2}\rangle|1\rangle_{1}|0\rangle_{2},|g_{1}g_{2}\rangle|0\rangle_{1}|1\rangle_{2}\big\}.
\end{eqnarray}
The $N$-excitation subspace ($N=2,3,4,...,\infty$) becomes:
\begin{eqnarray}\label{e6}
\Gamma_{N}&=&\big\{\Gamma_{a},\Gamma_{b},\Gamma_{c},\Gamma_{d}\big\},\cr
\Gamma_{a}&=&\big\{|e_{1}g_{2}\rangle|N-1\rangle_{1}|0\rangle_{2},|e_{1}g_{2}\rangle|N-2\rangle_{1}|1\rangle_{2},
..., \cr&&
 |e_{1}g_{2}\rangle|1\rangle_{1}|N-2\rangle_{2},|e_{1}g_{2}\rangle|0\rangle_{1}|N-1\rangle_{2}\big\},\cr
\Gamma_{b}&=&\big\{|g_{1}e_{2}\rangle|N-1\rangle_{1}|0\rangle_{2},|g_{1}e_{2}\rangle|N-2\rangle_{1}|1\rangle_{2},
..., \cr&&
 |g_{1}e_{2}\rangle|1\rangle_{1}|N-2\rangle_{2},|g_{1}e_{2}\rangle|0\rangle_{1}|N-1\rangle_{2}\big\},\cr
\Gamma_{c}&=&\big\{|g_{1}g_{2}\rangle|N\rangle_{1}|0\rangle_{2},|g_{1}g_{2}\rangle|N-1\rangle_{1}|1\rangle_{2},
..., \cr&&
 |g_{1}g_{2}\rangle|1\rangle_{1}|N-1\rangle_{2},|g_{1}g_{2}\rangle|0\rangle_{1}|N\rangle_{2}\big\},\cr
\Gamma_{d}&=&\big\{|e_{1}e_{2}\rangle|N-2\rangle_{1}|0\rangle_{2},|e_{1}e_{2}\rangle|N-3\rangle_{1}|1\rangle_{2},
..., \cr&&
 |e_{1}e_{2}\rangle|1\rangle_{1}|N-3\rangle_{2},|e_{1}e_{2}\rangle|0\rangle_{1}|N-2\rangle_{2}\big\}.
\end{eqnarray}

Suppose the density operator of the whole system is initially
$\rho(0)$. After an interaction time the density operator is given
by $\rho(t)$ $=$ $e^{-iHt}\rho(0)e^{iHt}$. Taking a partial trace
over two thermal fields, we can obtain the reduced density matrix
for two atoms which is spanned in the basis $\{$
$|e_{1}e_{2}\rangle$, $|e_{1}g_{2}\rangle$, $|g_{1}e_{2}\rangle$,
$|g_{1}g_{2}\rangle$ $\}$. Without any initial coherences from both
the atoms and fields, we calculate the reduced density matrix in
each excitation subspace, then sum over all the subspaces to get the
total reduced density matrix (7). The used Hilbert space is cut off
at $N\sim5(\bar{n}+1)$. For example, when $\bar{n}=0.1$, the Hilbert
space is cut off at $N=5$; when $\bar{n}=1$, the Hilbert space is
cut off at $N=15$; when $\bar{n}=10$, the Hilbert space is cut off
at $N=60$. Therefore, the density matrix $\rho_{a}$ has a simple
form:
\begin{eqnarray}\label{e7}
\rho_{a}&=&\left(\begin{array}{cccc}
A & 0 & 0 & G \\
0 & B & E & 0 \\
0 & E^{*} & C & 0 \\
G^{*} & 0 & 0 & D
\end{array}
\right).
\end{eqnarray}
The values of the elements in Eq. (7) depend on the initial state of
the whole system and they can be explicitly calculated by numerical
simulation. We adopt the Wootters's \cite{PRL-1998-80-2245}
concurrence $C(\rho_{a})$ to quantify the entanglement for two
atoms, which is defined as
\begin{eqnarray}\label{e8}
C(\rho_{a})=\max\bigg\{ 0,
\sqrt{\lambda_{1}}-\sqrt{\lambda_{2}}-\sqrt{\lambda_{3}}-\sqrt{\lambda_{4}}
\bigg\},
\end{eqnarray}
where $\lambda_{1}$, $\lambda_{2}$, $\lambda_{3}$ and $\lambda_{4}$
are the eigenvalues arranged in decreasing order of the following
matrix
\begin{eqnarray}\label{e9}
\xi&=&\rho_{a}(\sigma_{y}\otimes\sigma_{y})\rho_{a}^{*}(\sigma_{y}\otimes\sigma_{y}),
\end{eqnarray}
where $\rho_{a}^{*}$ represents the complex conjugation of
$\rho_{a}$ and $\sigma_{y}$ is the corresponding Pauli matrix.
Therefore the concurrence for the two atoms takes the form:
\begin{eqnarray}\label{e10}
C(\rho_{a})=2\max\bigg\{ 0,|E|-\sqrt{AD},|G|-\sqrt{BC}\bigg\}.
\end{eqnarray}
It has been shown that using the concurrence to deal with the mixed
state of any two qubits is an easy evaluation of entanglement of
formation, and it satisfies the following important conditions of
the measure of entanglement: (1) $C(\rho_a)=0$ if and only if
$\rho_a$ is separable; (2) For all local unitary transformations,
$C(\rho_a)$ keeps invariant; (3) $C(\rho_a)$ does not increase under
local general measurements, classical communications, and
post-selection of subensemble. For unentangled atoms
$C(\rho_{a})=0$, while $C(\rho_{a})=1$ for the maximally entangled
state. Although the exact analytic expression for Eq. (10) is very
complicated as the excitation number becomes infinite, there are
several limiting situations where the whole dynamics can be
simplified. Under these limiting conditions, we can obtain an
effective Hamiltonian that reflects the essentially physical
processes and provides a basic explanation for the numerical
results.

\section{Generation of Atomic Entanglement}

Assume that the initial state for the two atoms is
$|e_{1}g_{2}\rangle$. Then the density matrix of the whole system is
\begin{eqnarray}\label{e11}
\rho (0)&=&|e_{1}g_{2}\rangle\langle e_{1}g_{2}| \otimes\sum_{n_{1},
n_{2}=0}^{\infty} P_{1}(n_{1})P_{2}(n_{2})|n_{1}n_{2}\rangle\langle
n_{1}n_{2}|.
\end{eqnarray}
Set $\bar{n}_{1}$ $=$ $\bar{n}_{2}$ $=$ $\bar{n}$ $=$ $1$, the
concurrence is plotted for the cases of large frequency detuning and
large hopping strength in Fig. 2 and Fig. 3, respectively. In order
to explain the underlying physics explicitly, we introduce two
delocalized bosonic modes $b_{1}$ and $b_{2}$, which are defined as
$b_{1}$ $=$ $(a_{1}+a_{2})/\sqrt{2}$ and $b_{2}$ $=$
$(a_{1}-a_{2})/\sqrt{2}$. We first consider the symmetric coupling
situation, i.e., $g_{1}$ $=$ $g_{2}$ $=$ $g$, and set $\delta_{1}$
$=$ $\delta_{2}$ $=$ $\delta$ for simplicity. In terms of the new
operators, the Hamiltonian $H$ can be rewritten as follows:
\begin{eqnarray}\label{e12}
H&=&\delta_{1}^{'}b_{1}^{\dagger}b_{1}+\delta_{2}^{'}b_{2}^{\dagger}b_{2}\cr&&
+\frac{g}{\sqrt{2}}\bigg[b_{1}(S_{1}^{+}+S_{2}^{+})+b_{2}(S_{1}^{+}-S_{2}^{+})+H.c.\bigg],
\end{eqnarray}
where $\delta_{1}^{'}$ $=$ $(\delta+J)$ and $\delta_{2}^{'}$ $=$
$(\delta-J)$, which indicates that the frequencies of two
delocalized field modes are shifted from the cavity frequency due to
coherent photon hopping. In the interaction picture with respect to
$H_{0}$ $=$
$\delta_{1}^{'}b_{1}^{\dagger}b_{1}+\delta_{2}^{'}b_{2}^{\dagger}b_{2}$,
we get the atom-field interaction Hamiltonian
\begin{eqnarray}\label{e13}
&H_{I}=\frac{g}{\sqrt{2}}\bigg[e^{-i\delta_{1}^{'}t}b_{1}(S_{1}^{+}
+S_{2}^{+})+e^{-i\delta_{2}^{'}t}b_{2}(S_{1}^{+}-S_{2}^{+})+H.c.\bigg].
\end{eqnarray}
Under the conditions $\delta_{1}^{'}$ $\gg$
$\sqrt{\bar{n}_{1}+1}g/\sqrt{2}$ and  $\delta_{2}^{'}$ $\gg$
$\sqrt{\bar{n}_{2}+1}g/\sqrt{2}$, there is no energy exchange
between the atomic system and the field modes, and the two atoms are
coupled to each other via exchange of virtual photons. Then the
effective Hamiltonian is given by
\cite{PRA2008-78-063805,PRL-2000-85-2392}:
\begin{eqnarray}\label{e14}
H_{eff}&=&-\sum_{i=1}^{2}\big[(\frac{g^2}{2\delta_{1}^{'}}b_{1}^{\dagger}b_{1}+\frac{g^2}{2\delta_{2}^{'}}b_{2}^{\dagger}b_{2})
(|e_{i}\rangle\langle e_{i}|-|g_{i}\rangle\langle
g_{i}|)+\lambda|e_{i}\rangle\langle e_{i}|\big]+\lambda^{'}(
S_{1}^{+}S_{2}^{-}+H.c.),\cr&&
\end{eqnarray}
where $\lambda$ $=$
$\frac{g^2}{2\delta_{1}^{'}}+\frac{g^2}{2\delta_{2}^{'}}$ and
$\lambda^{'}$ $=$
$\frac{g^2}{2\delta_{1}^{'}}-\frac{g^2}{2\delta_{2}^{'}}$. The
evolution of the atomic system is $\cos(\lambda^{'}
t)|e_{1}g_{2}\rangle$ $-$ $i\sin(\lambda^{'} t)|g_{1}e_{2}\rangle$,
which is independent of the field states. We here have discarded the
trivial common phase factor $e^{-i\lambda^{'} t}$. Therefore, for
large photon hopping strength the local resonant thermal fields can
induce maximal atomic entanglement, which is distinguished from the
case for two atoms interacting commonly with one single-mode thermal
field \cite{PRA-65-040101-2002}. The effective Hamiltonian (14) is
valid under the condition that the difference between $\delta$ and
$J$ is much larger than $g$. When $\delta\sim J$, the eigenfrequency
of the normal delocalized mode $b_{2}^{'}$ approaches the atomic
transition frequency so that the atoms can exchange energy with this
delocalized mode and the effective Hamiltonian (14) is invalid. This
leads to the atom-cavity entanglement and deteriorates the atom-atom
entanglement, which accounts for the appearance of the dips in Fig.
2 and Fig. 3 for $\delta$ near $J$. In the regime $\delta\sim J$,
the atom-cavity system undergoes Rabi oscillations, and each atom is
not only entangled with each other but also with the cavity modes.
One can maximize the atom-atom entanglement by choosing the optimal
ratio of $g$ to $\delta^{'}_{2}$ and $t$. The entanglement is very
sensitive to the ratio $\delta^{'}_{2}/g$. This accounts for the
maxima and abrupt changes of the concurrence inside the dips of Fig.
2 and Fig. 3. Due to the spread of the Rabi frequencies
corresponding to different photon numbers, the atoms cannot be
completely disentangled with the cavity modes and the maximal
atom-atom entanglement cannot be obtained. In fact, the physics
discussed here is similar to that discussed in Ref.
\cite{PRL-2000-85-2392} which proposed the generation of maximally
entangled states for two atoms via dispersive interaction with a
single cavity mode.

When two atoms are both initially in their excited states
$|e_{1}e_{2}\rangle$, we find that the atomic entanglement could
also be induced by two thermal fields with appropriate parameters,
as shown in Fig. 4. In the case of Fig. 4(a), the atomic
entanglement arises from either atom's emission of one photon into
the same cavity modes. When the probability for this event is zero,
the entanglement vanishes. In Fig. 5, we plot the expectation value
of total atomic excitation number $\langle \sigma_{z1}+\sigma_{z2}
\rangle$ versus $gt$ for the same parameters as Fig. 4(a). The
result shows the peak of concurrence coincides with the case when
$\langle \sigma_{z1}+\sigma_{z2} \rangle$ is nearest to zero, which
corresponds to the regime where probability for either atom's
emission of one photon is maximized. Interestingly, for symmetric
atom-cavity couplings the entanglement exhibits periodical sudden
birth and death when the atomic transition is slightly detuned from
at least one delocalized mode, as shown in Fig. 4(a). Note that the
atoms start from a product state, which is distinguished from the
previous studies of entanglement sudden birth and death in which the
atoms start from an entangled state
\cite{JPB-39-S621-2006,PRA-77-054301-2008,PRL-101-080503-2008}. The
duration of the entanglement death is much longer than that of the
entanglement life. Unlike previous examples, the peaks of the
concurrence do not decrease monotonically as the interaction time
increases. The entanglement dynamics for the initial state
$|g_{1}g_{2}\rangle$ is similar with that for initial state
$|e_{1}e_{2}\rangle$, as shown in Fig. 4(b). The entanglement
dynamics for both the initial states $|e_1e_2\rangle$ and
$|g_1g_2\rangle$ is similar to the case when the two atoms interact
with a common cavity mode
\cite{PRA-65-040101-2002,OC-283-4676-2010}. On the other hand, for
asymmetric atom-cavity couplings, i.e., $g_{1}$ $\neq$ $g_{2}$, the
atomic entanglement can be generated when the atoms resonantly
interact with two thermal fields, as shown in Fig. 4(c). The
vanishes of entanglement in Fig. 4 corresponds to entanglement
sudden death that also appears in the system with two atoms
independently interact with the respective cavities
\cite{JPB-39-S621-2006}. The considered system involves four
subsystems so that the whole system evolution is very complex and
does not exhibit periodical features. This accounts for the fact
that distributions of peaks and dips in Fig. 3 and Fig. 4 are
somewhat random.

\section{Freeze of Atomic Entanglement}

Moving forward, we also investigate the system evolution when the
atoms are initially in the maximally entangled state
$|\Psi_{m}\rangle$ $=$
$(|e_{1}g_{2}\rangle+|g_{1}e_{2}\rangle)/\sqrt{2}$. In the resonant
interaction situation, as shown in Fig. 5, for $J\gg g$, the
entanglement can be frozen. This is due to the fact in this case the
conditions $\delta_{1}^{'}$ $\gg$ $\sqrt{\bar{n}_{1}+1}g/\sqrt{2}$
and $\delta_{2}^{'}$ $\gg$ $\sqrt{\bar{n}_{2}+1}g/\sqrt{2}$ are
satisfied, so that the effective Hamiltonian of Eq. (14) dominates
the system evolution and the maximally entangled state
$|\Psi_{m}\rangle$ becomes an eigenstate of this effective
Hamiltonian. As the values of $J/g$ decreases, the concurrence
decreases. Especially when $J/g$ $\rightarrow$ $0$, the sudden death
of entanglement emerges, since in this case each atom independently
interacts with a thermal field, which causes the leakage of the
atomic coherence into the fields.

The influence of temperature, i.e., mean photon numbers $\bar{n}$,
on the atom-atom entanglement are taken into consideration, as
plotted in Fig. 7. The result shows that the larger the mean photon
number, the smaller the concurrence. This phenomenon can be
understood in the following way. The probability that the atoms
exchange energy with the fields increases with mean photon numbers.
When the photon numbers are large, the conditions $\delta_{1}^{'}$
$\gg$ $\sqrt{\bar{n}_{1}+1}g/\sqrt{2}$ and $\delta_{2}^{'}$ $\gg$
$\sqrt{\bar{n}_{2}+1}g/\sqrt{2}$ are not satisfied and the effective
Hamiltonian (14) is not valid. When the photon number is
sufficiently small so that the effective Hamiltonian (14) is valid,
the effective atom-atom coupling strength is
$\lambda^{'}=g^{2}/J=0.1g$, which is photon-number-independent. When
$\lambda^{'}t=\pi/4$, i.e., $gt=2.5\pi$, the excitation is
approximately equally shared by the two atoms and the concurrence
reaches the maxima. At the time $t=\pi/(2\lambda^{'})=5\pi$, the
excitation is completely transferred from the first atom to the
second one. So that the concurrence vanishes.

\section{Atomic Entanglement Dynamics with Dissipation Being Included}
In all the above discussions, we have just supposed that the entire
system is ideally isolated from the outside environment. Taking
cavity decay and atomic spontaneous emission into account, the
master equation for the density matrix $\rho(t)$ of the system can
be expressed as:
\begin{eqnarray}\label{e15}
\dot{\rho}(t)&=&-i\big[H,\rho(t)\big]\cr&&+\frac{\kappa}{2}
(\bar{n}+1)\sum_{i=1}^{2}\big[2a_{i}\rho(t)
a_{i}^{\dagger}-a_{i}^{\dagger}a_{i}\rho(t)-\rho(t)
a_{i}^{\dagger}a_{i}
\big]\cr&&+\frac{\kappa}{2}\bar{n}\sum_{i=1}^{2}\big[2a_{i}^{\dagger}\rho(t)
a_{i}-a_{i}a_{i}^{\dagger}\rho(t)-\rho(t) a_{i}a_{i}^{\dagger}
\big]\cr&&+\frac{\gamma}{2}(\bar{n}+1)\sum_{i=1}^{2}\big[2S_{i}^{-}\rho(t)
S_{i}^{+}-S_{i}^{+}S_{i}^{-}\rho(t)-\rho(t) S_{i}^{+}S_{i}^{-}
\big]\cr&&+\frac{\gamma}{2}\bar{n}\sum_{i=1}^{2}\big[2S_{i}^{+}\rho(t)
S_{i}^{-}-S_{i}^{-}S_{i}^{+}\rho(t)-\rho(t) S_{i}^{-}S_{i}^{+}
\big],
\end{eqnarray}
where $\kappa$ and $\gamma$ denote the cavity decay rate and the
atomic spontaneous emission rate, respectively. The dependence of
the concurrence on the cooperative parameter
$C_{\kappa\gamma}=g^2/(\kappa\gamma)$ is plotted, as shown in Fig.
8. When the detunings between atomic transition and the delocalized
modes are much larger than the atom-cavity coupling strength, the
entanglement dynamics is much more sensitive to the atomic decay
than to the cavity loss as in this case the cavity modes are only
virtually excited and they are decoupled from the atomic
entanglement dynamics. Therefore, the scheme is favorable when
$\gamma$ $\ll$ $\kappa$. Here we choose $\gamma$ $=$ $0.1\kappa$ in
Fig 8(a) and (c), and $\gamma$ $=$ $\kappa$ in Fig 8(b) and (d). The
difference between Fig 8(a)((c)) and Fig 8(b)((d)) is that the
concurrence decreases quickly as the atomic decay increases, which
demonstrates the cavity decay is not the dominant factor
deteriorating the entanglement.

\section{Conclusions}

In conclusion, we have studied the entanglement dynamics of two
identical atoms interacting with two single-mode thermal fields
based on the coupled-cavity system. The atom-atom concurrence is
calculated under various conditions. The results show that the
entanglement behaviors depend upon the photon hopping strength, the
atom-cavity detuning, and the initial atomic state. It is
demonstrated that under the situation where each atom is resonant
with the local field mode, the maximal entanglement for two atoms
can be generated with one atom in the ground state and the other in
the excited state initially, which is impossible for the case with
two atoms interacting with a single thermal cavity mode. When the
two atoms are both initially in the excited state, the entanglement
sudden birth and death occur periodically, and the peaks of the
concurrence do not decrease monotonically when the evolution time
increases. Furthermore, we show that the initial two-atom maximal
entanglement can be frozen with suitable choice of the parameters.
The effect of dissipation on the entanglement is analyzed.

L.T.S., H.Z.W., X.Y.C. and S.B.Z. acknowledge support from the Major
State Basic Research Development Program of China under Grant No.
2012CB921601, National Natural Science Foundation of China under
Grant No. 10974028, the Doctoral Foundation of the Ministry of
Education of China under Grant No. 20093514110009, and the Natural
Science Foundation of Fujian Province under Grant No. 2009J06002.
Z.B.Y acknowledges support from the National Basic Research Program
of China under Grants No. 2011CB921200 and No. 2011CBA00200, and the
China Postdoctoral Science Foundation under Grant No. 20110490828.

\clearpage

\section*{List of Figure Captions}

Fig. 1. (Color online) The proposed experimental setup. Two
two-level atoms are respectively trapped in two coupled cavities,
each of which is initially in a thermal state.

\noindent Fig. 2. (Color online)  Atom-atom concurrence as a
function of the evolution time and photon hopping strength with
$\bar{n}$ $=$ $1$ and $\delta=10g$. The atoms are initially in the
state $|e_{1}g_{2}\rangle$.

\noindent Fig. 3. (Color online)  Atom-atom concurrence as a
function of the evolution time and photon hopping strength with
$\bar{n}$ $=$ $1$ and $J=25g$. The atoms are initially in the state
$|e_{1}g_{2}\rangle$.

\noindent Fig.4. (Color online)  Atom-atom concurrence as a function
of the evolution time and hopping strength when $\bar{n}$ $=$ $0.1$
: (a) $g_{1}$ $=$ $g_{2}$ $=$ $g$ for the initial state
$|e_{1}e_{2}\rangle$; (b) $g_{1}$ $=$ $g_{2}$ $=$ $g$ for the
initial state $|g_{1}g_{2}\rangle$; (c) $\delta$ $=$ $0$, $g_{1}$
$=$ $J$ and $g_{2}$ $=$ $g$ for the initial state
$|e_{1}e_{2}\rangle$.

\noindent Fig.5. (Color online) Dash red line represents the atomic
population inversion $\langle \sigma_{z1}+\sigma_{z2} \rangle $ and
solid black line represents the average photon number $\langle
a_{1}^{\dagger} a_{1} \rangle $, which are as a function of the
evolution time when $\bar{n}$ $=$ $0.1$, $J=20g$ and $\delta=18.5g$
for the initial state $|e_{1}e_{2}\rangle$.

\noindent Fig.6. (Color online)  Atom-atom concurrence as a function
of evolution time and hopping strength with $\delta$ $=$ $0$ and
$\bar{n}$ $=$ $0.1$. The atoms are initially in the entangled state
$(|e_{1}g_{2}\rangle+|g_{1}e_{2}\rangle)/\sqrt{2}$.

\noindent Fig.7. (Color online) Atom-atom concurrence as a function
of evolution time with $\delta$ $=$ $0$, $J$ $=$ $10g$. The atoms
are initially in: (a) $|e_{1}g_{2}\rangle$; (b)
$(|e_{1}g_{2}\rangle+|g_{1}e_{2}\rangle)/\sqrt{2}$.

\noindent Fig.8. (Color online) Atom-atom concurrence as a function
of the evolution time $gt$ and cooperative parameter
$C_{\kappa\gamma}$ when $\bar{n}$ $=$ $0.1$. The atoms are in the
initial state $|e_{1}g_{2}\rangle$ with $\delta=0$ and $J=10g$: (a)
$\gamma=0.1\kappa$; (b) $\gamma=\kappa$. While the atoms are in the
initial state $(|e_{1}g_{2}\rangle+|g_{1}e_{2}\rangle)/\sqrt{2}$
with $\delta=15g$ and $J=5g$: (c) $\gamma=0.1\kappa$; (d)
$\gamma=\kappa$.

\clearpage

\begin{figure}
\centering
\includegraphics[width=0.5\columnwidth]{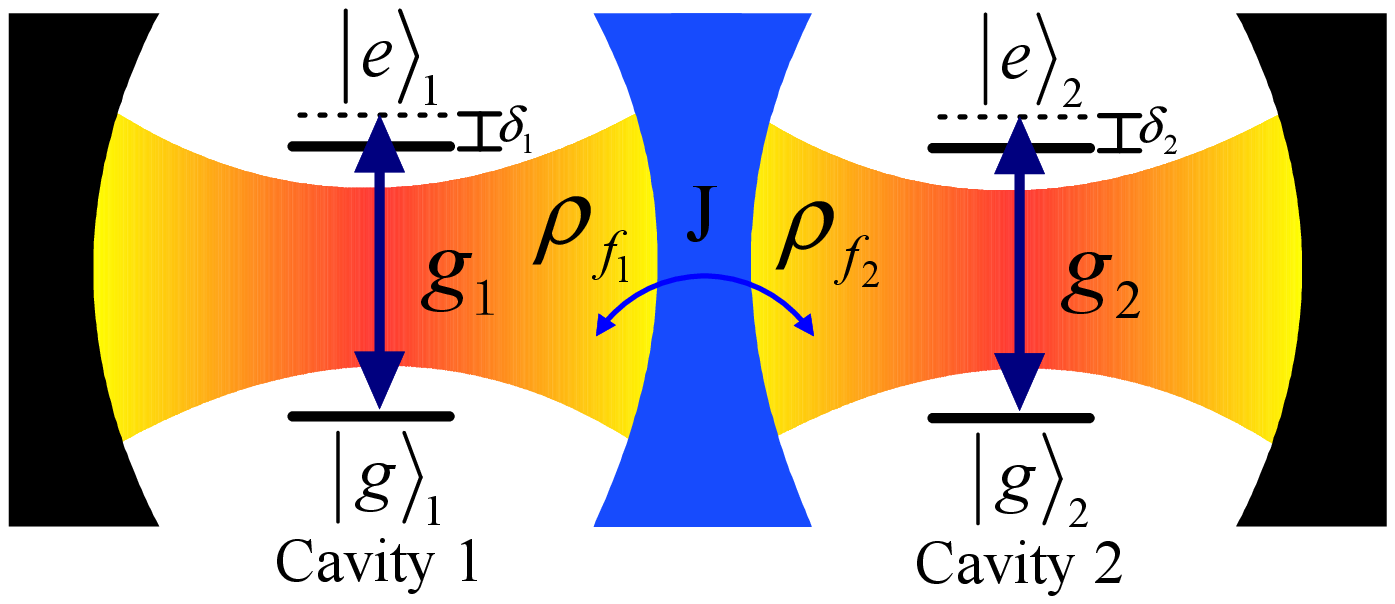} \caption{}
\end{figure}

\begin{figure}
\centering
\includegraphics[width=0.8\columnwidth]{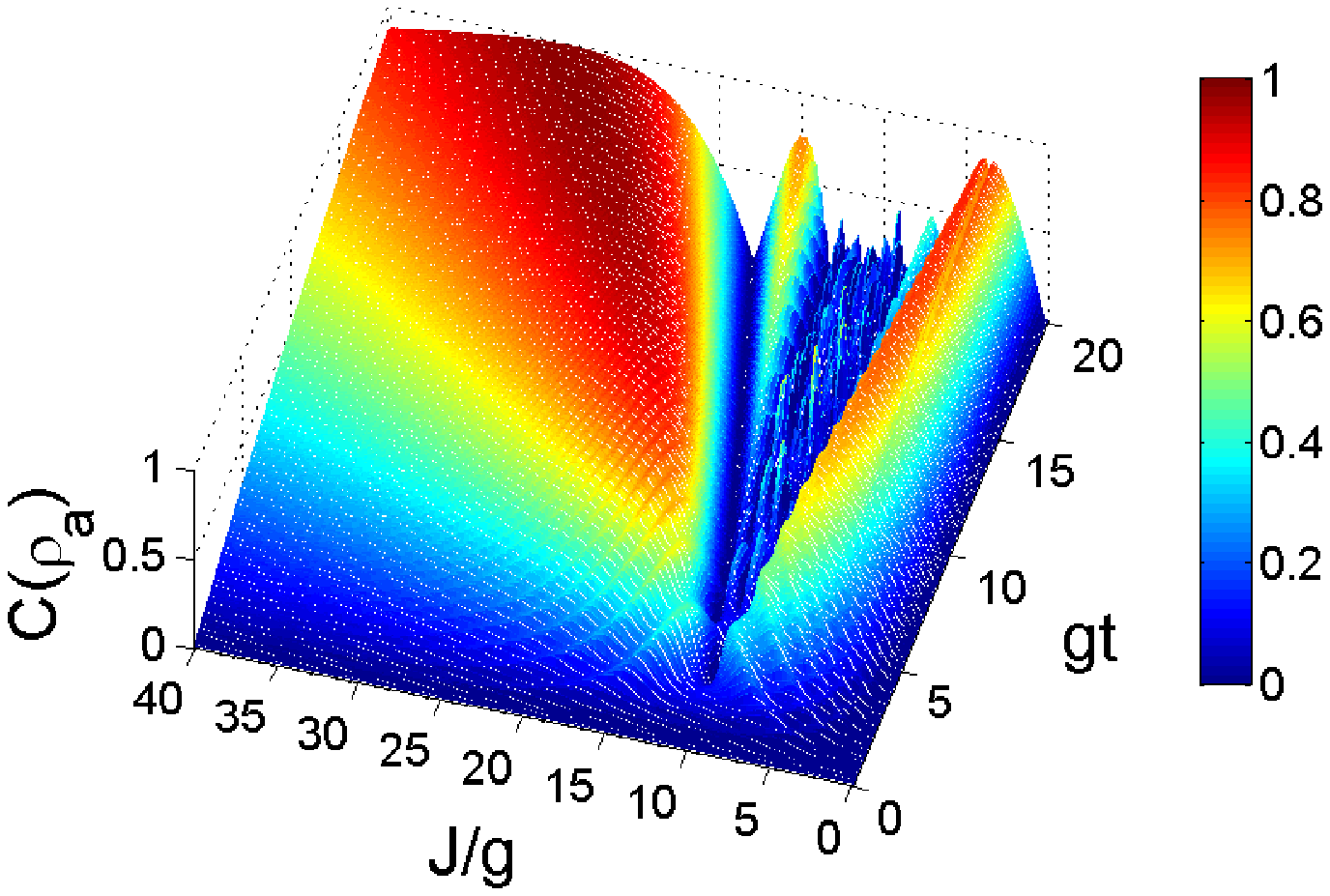} \caption{}
\end{figure}

\begin{figure}
\centering
\includegraphics[width=0.8\columnwidth]{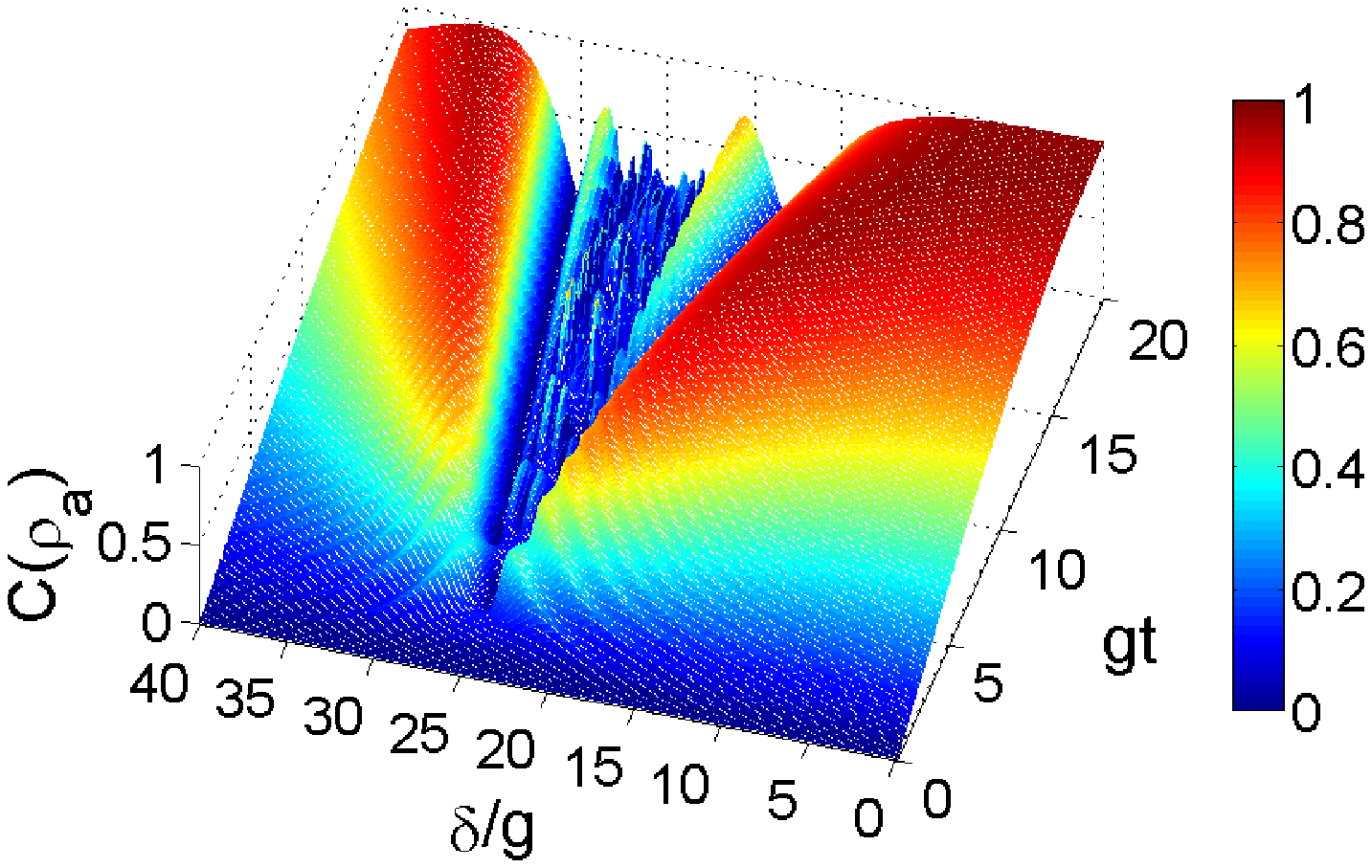} \caption{}
\end{figure}

\begin{figure}
 \centering \subfigure[]{
\label{Fig.sub.a}
\includegraphics[width=0.6\columnwidth]{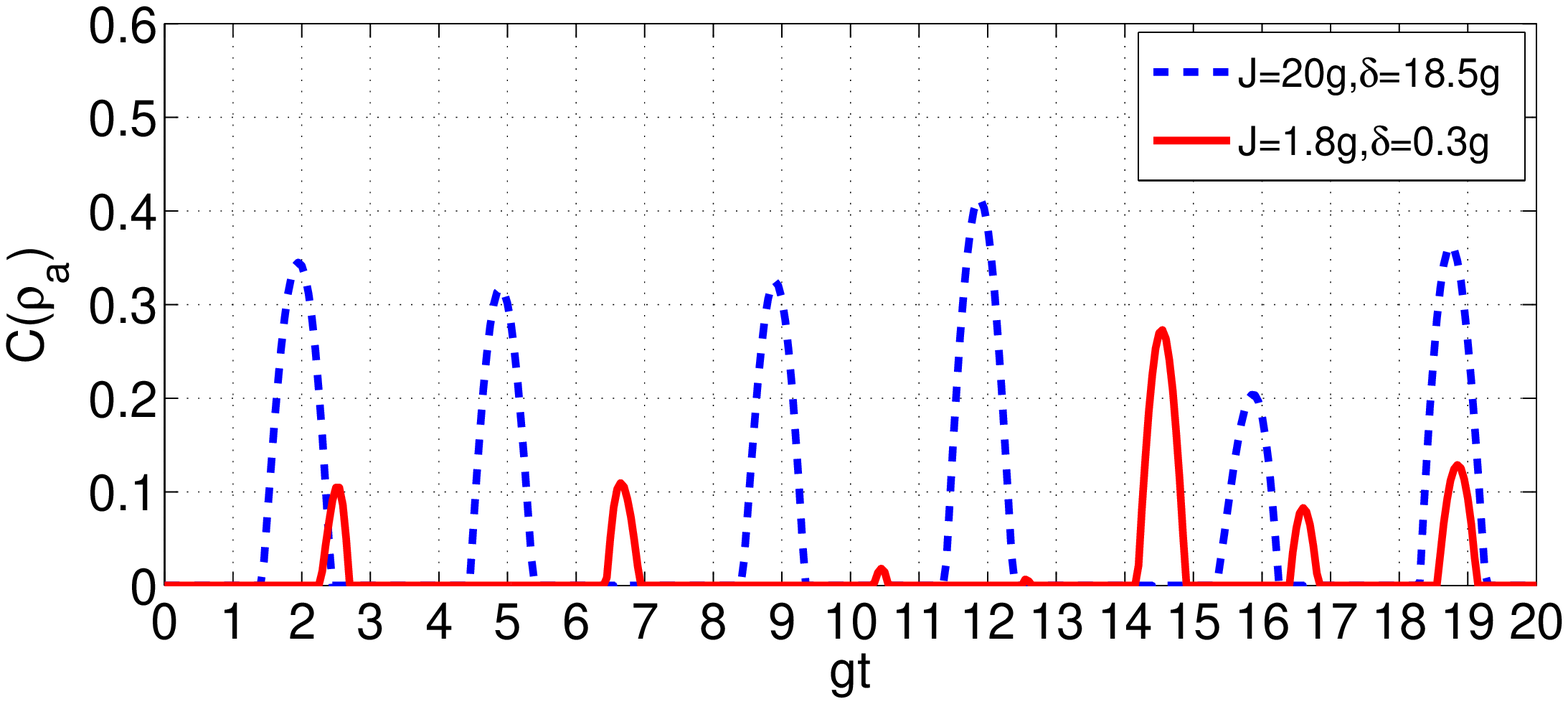}}
\subfigure[]{ \label{Fig.sub.c}
\includegraphics[width=0.6\textwidth]{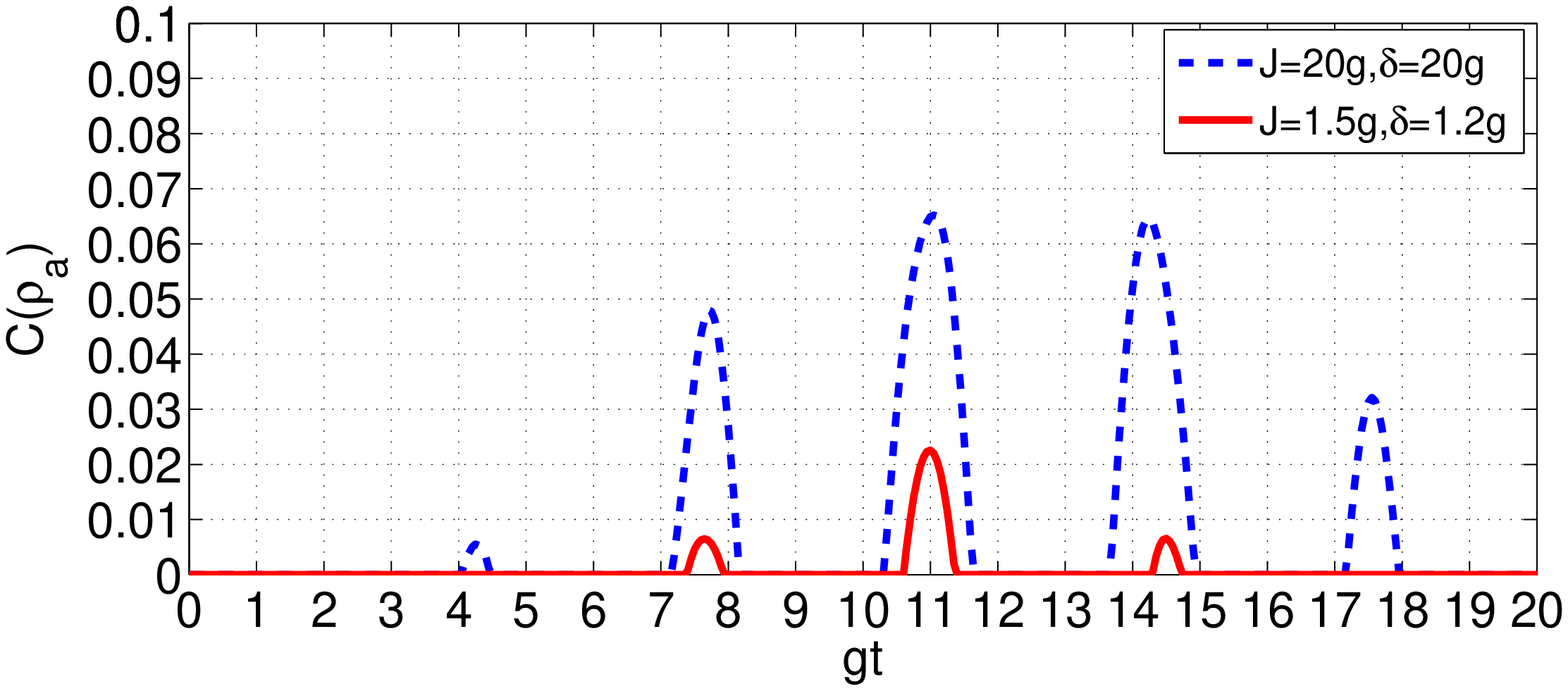}}
\subfigure[]{ \label{Fig.sub.d}
\includegraphics[width=0.7\columnwidth]{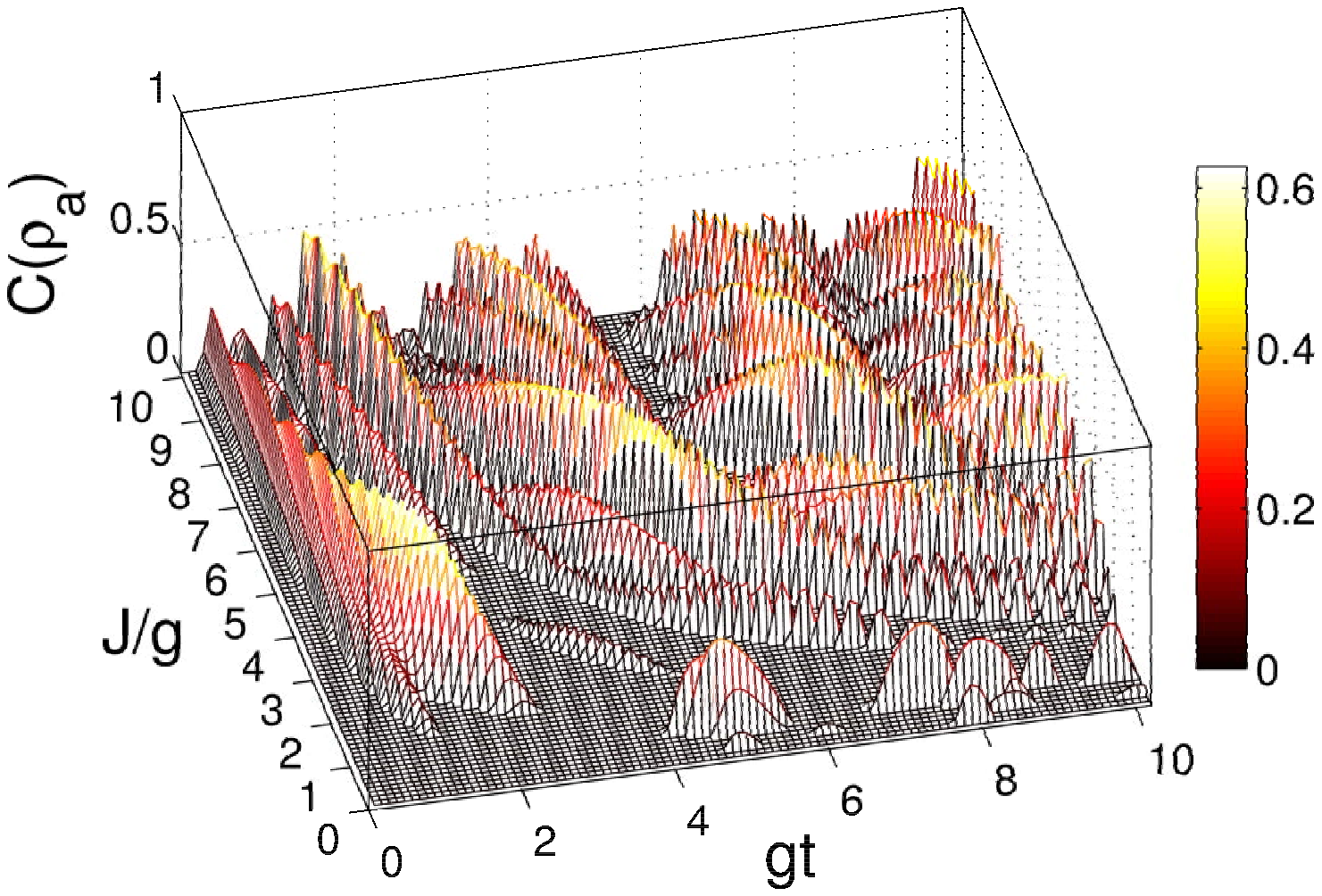}} \caption{}
\end{figure}

\begin{figure}
\centering
\includegraphics[width=0.6\columnwidth]{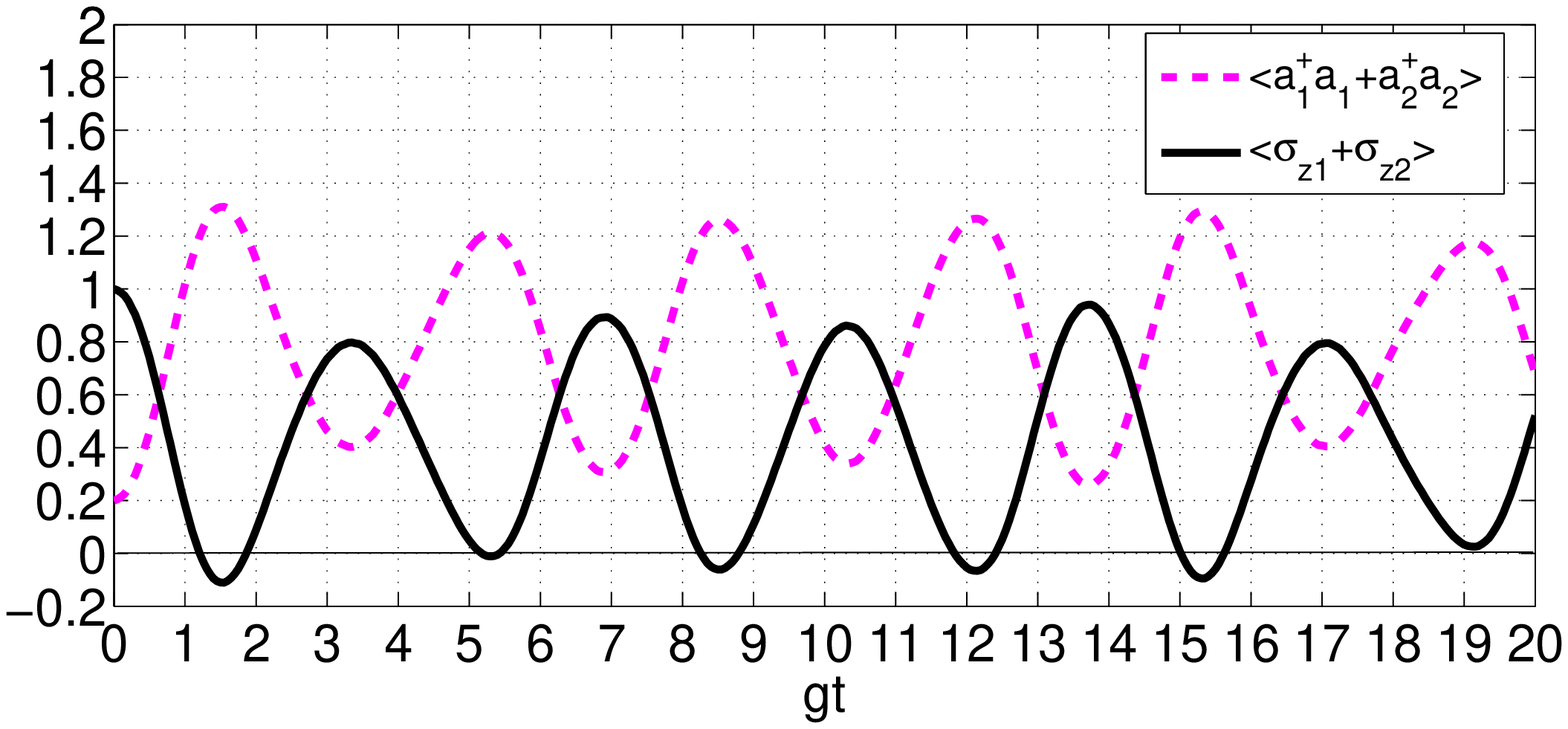} \caption{}
\end{figure}

\begin{figure}
\centering
\includegraphics[width=0.8\columnwidth]{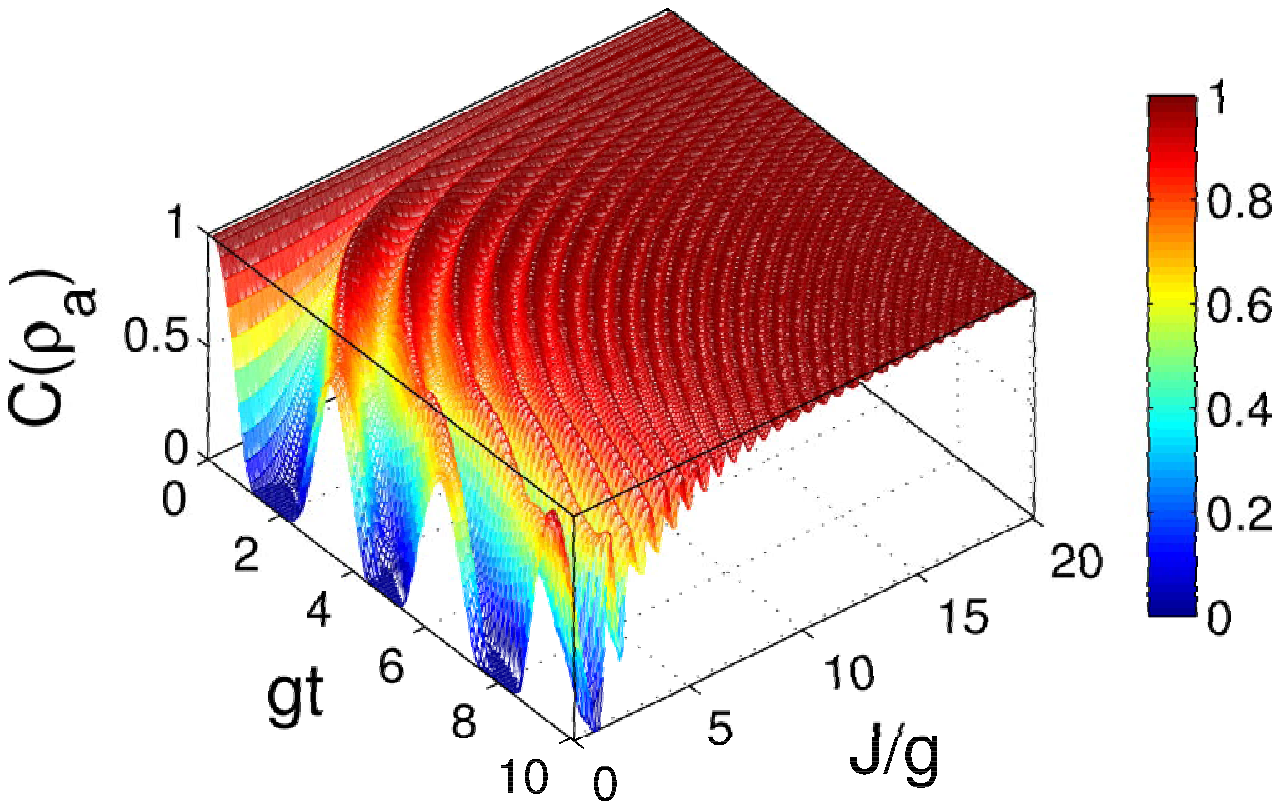} \caption{}
\end{figure}

\begin{figure}
\centering \subfigure[]{ \label{Fig.sub.a}
\includegraphics[width=0.5\columnwidth]{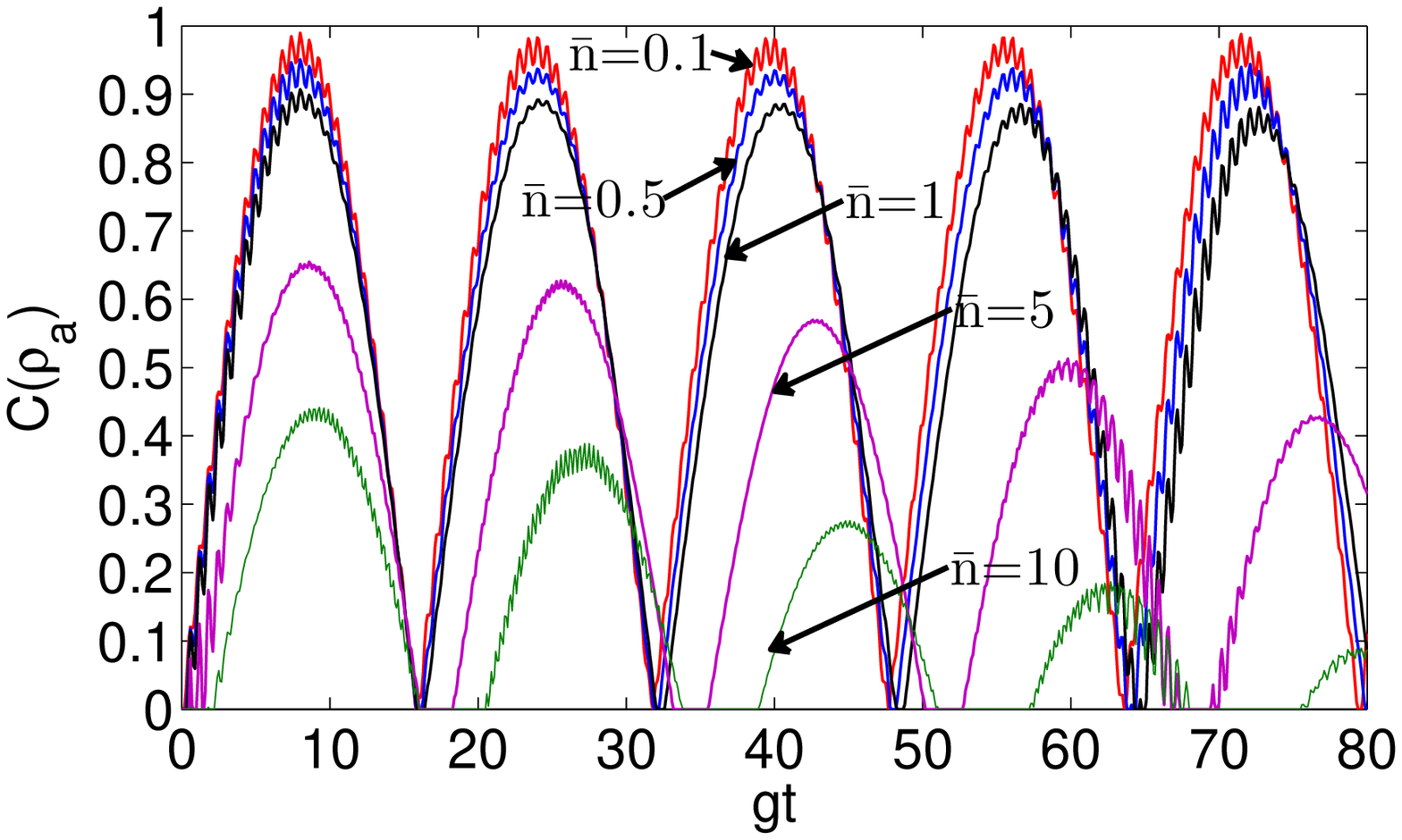}}
\subfigure[]{ \label{Fig.sub.b}
\includegraphics[width=0.5\columnwidth]{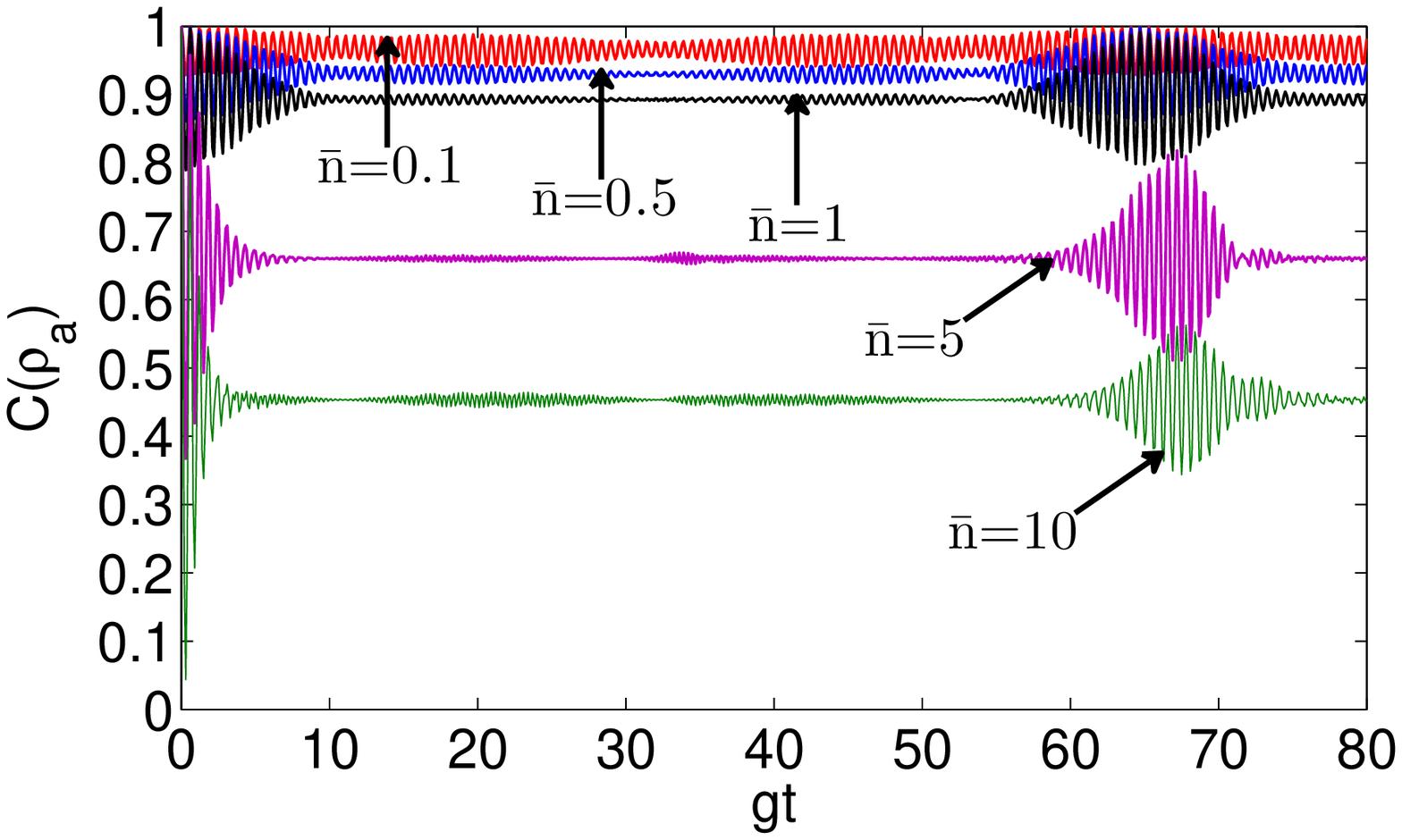}}
\caption{}
\end{figure}

%ebb fig7egv2.pdf
\begin{figure}
\centering
\includegraphics[width=0.45\columnwidth]{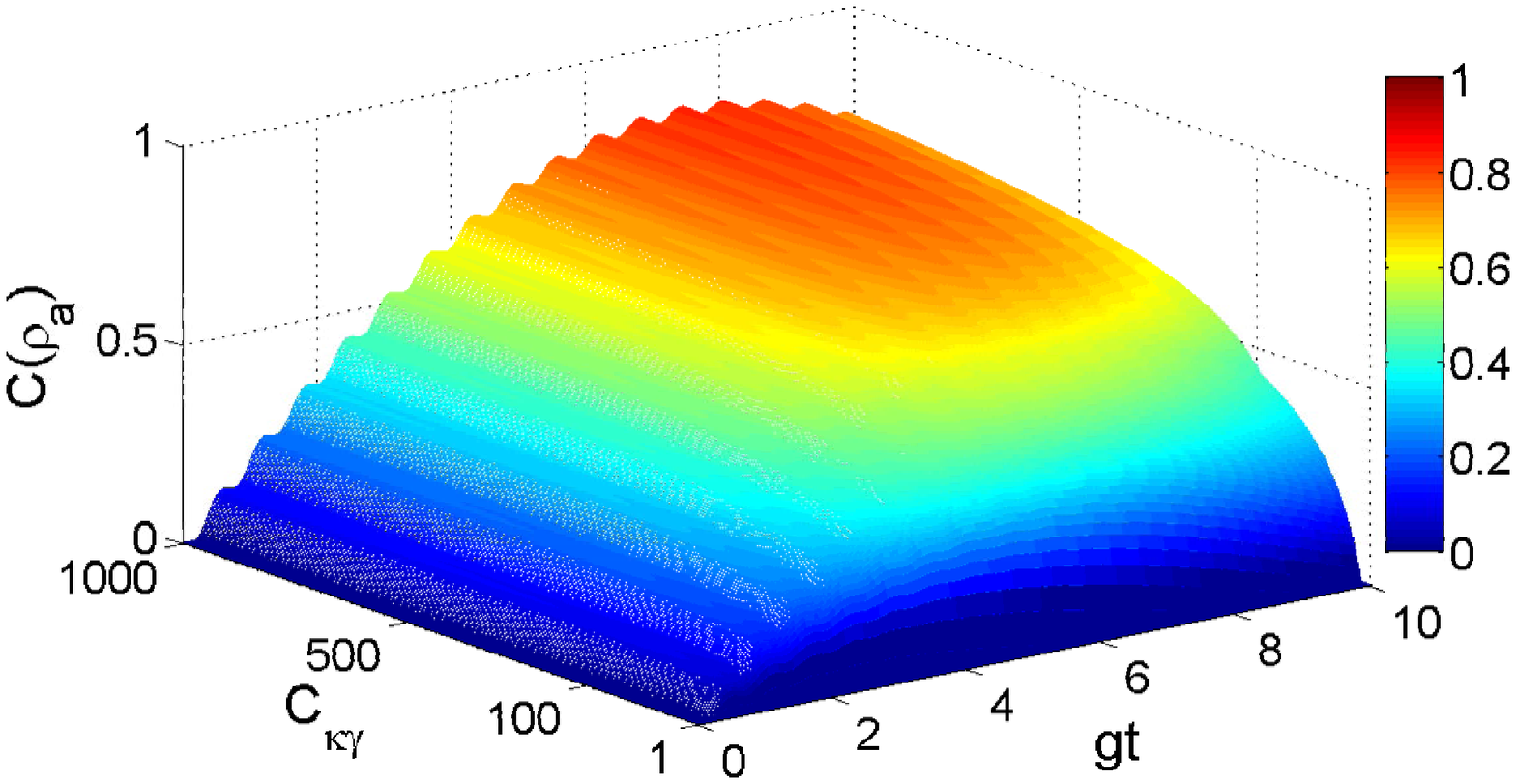}(a)
\includegraphics[width=0.45\columnwidth]{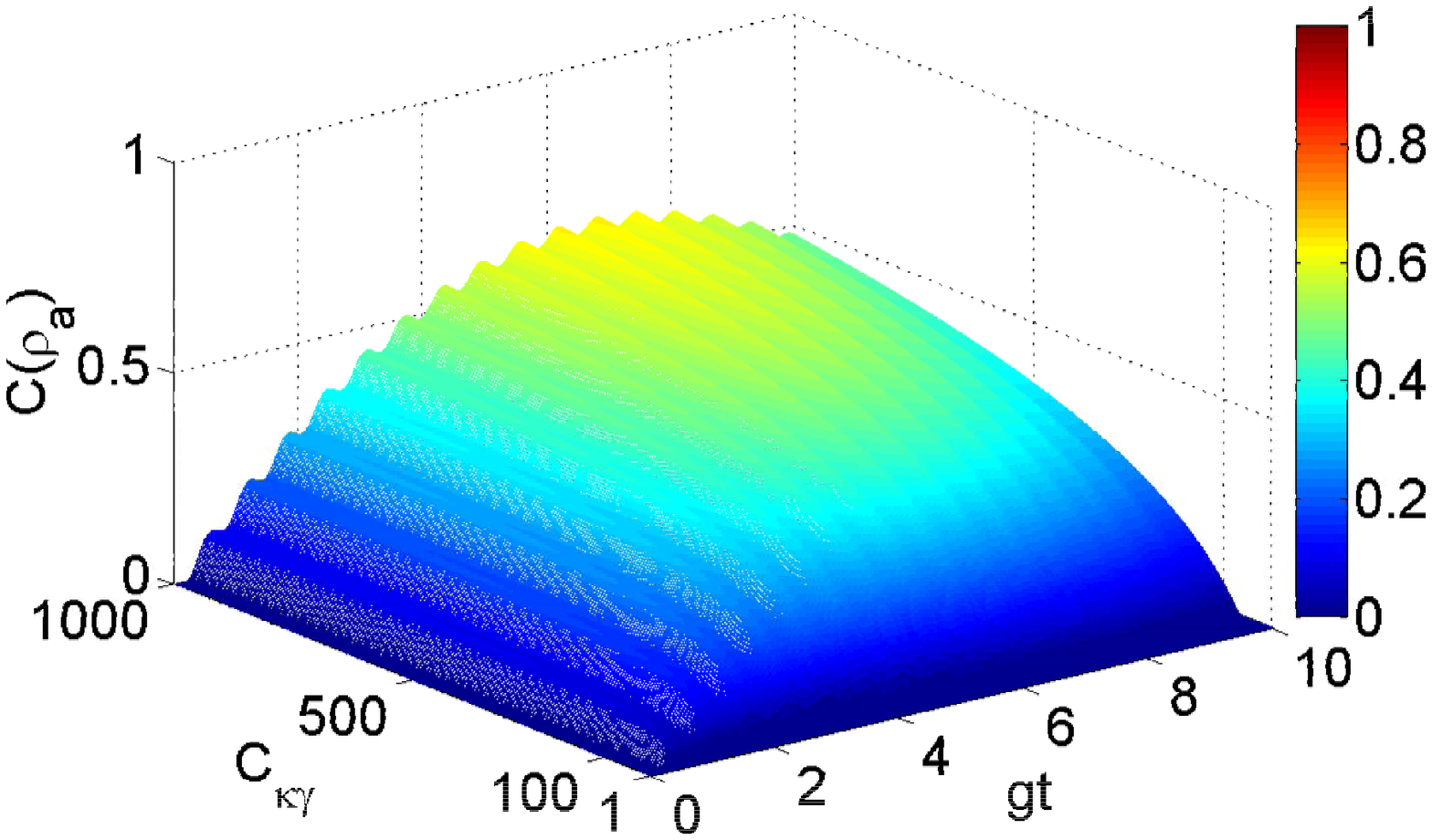}(b)
\includegraphics[width=0.45\columnwidth]{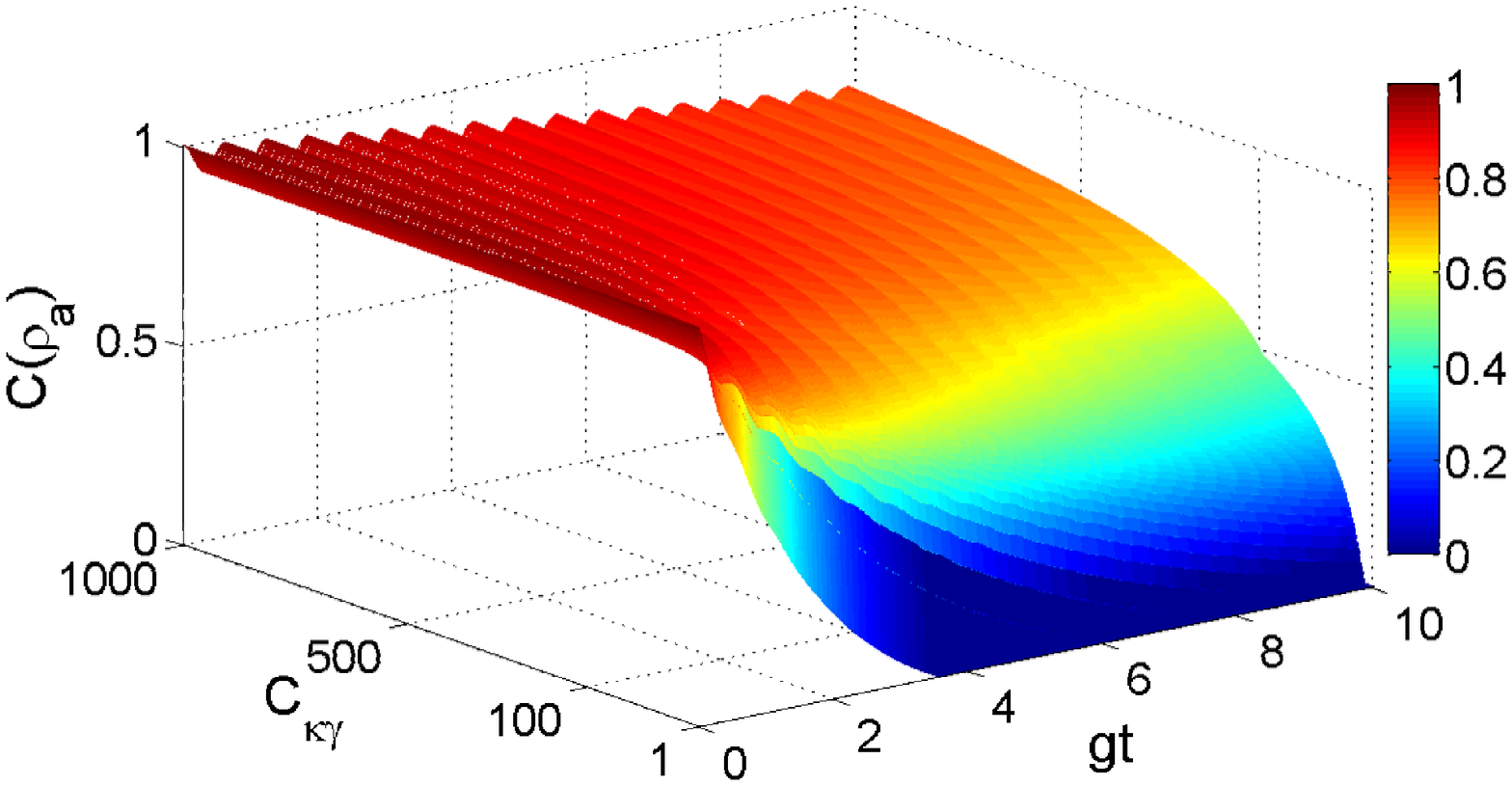}(c)
\includegraphics[width=0.45\columnwidth]{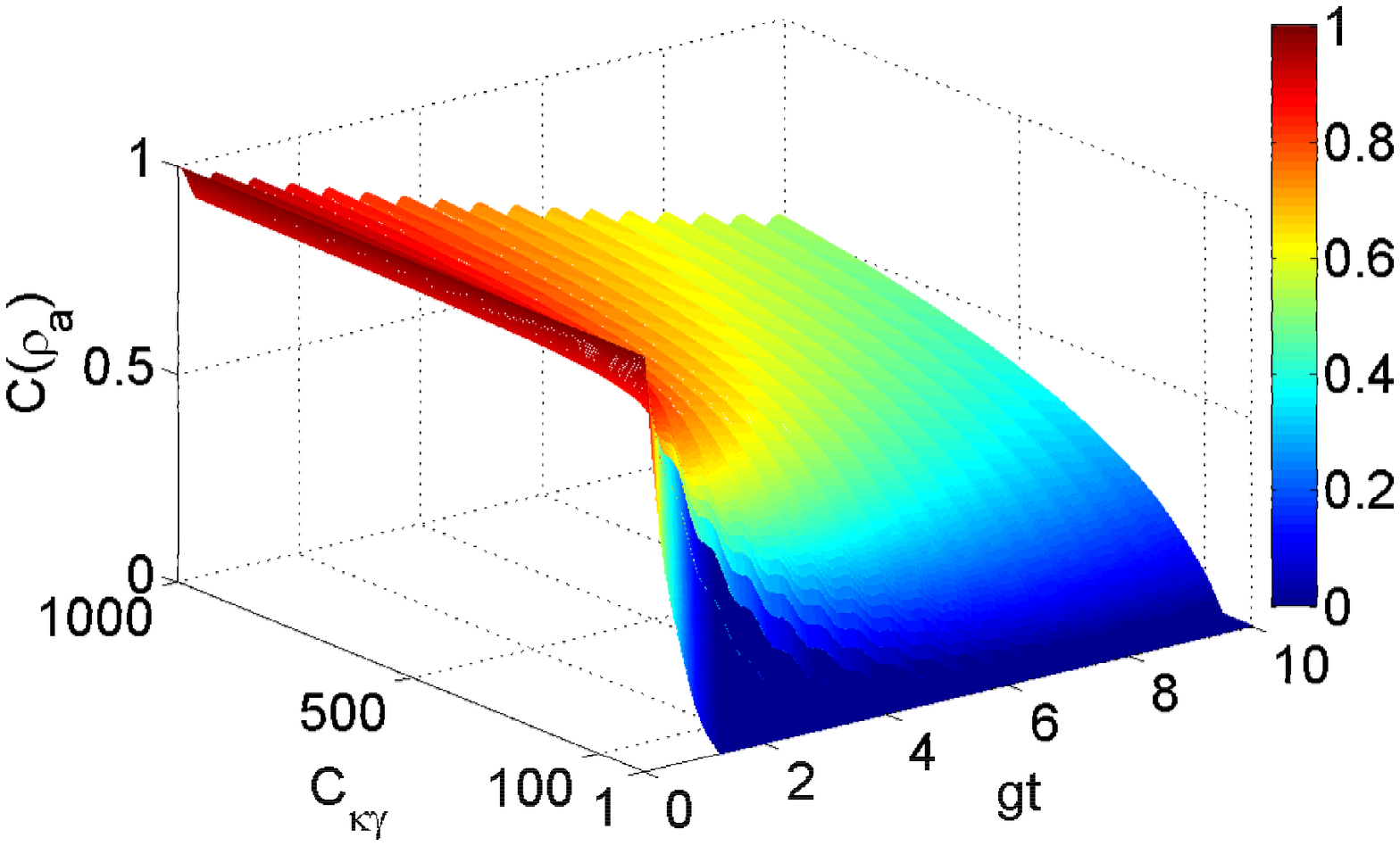}(d)
\caption{}
\end{figure}

\end{document}